\documentclass{emulateapj}
\usepackage{apjfonts}
\usepackage{epsfig}
\usepackage{natbib}

\newcommand{\Msun}{M_{\sun}}

\shorttitle{A Realistic Disk Galaxy}
\shortauthors{Governato et al.}

\begin{document}

\title{The Formation of a Realistic Disk Galaxy  in  $\Lambda$ dominated 
cosmologies }
\author{F.Governato\altaffilmark{1,2,9},
        L. Mayer\altaffilmark{3}, 
        J.Wadsley\altaffilmark{4}, 
        J.P.Gardner\altaffilmark{5},
        B.Willman\altaffilmark{6},
        E.Hayashi\altaffilmark{7}
        T. Quinn\altaffilmark{1},
        J.Stadel\altaffilmark{2} and
        G.Lake\altaffilmark{8}} 

\altaffiltext{1}{Department of Astronomy, Univ. of Washington,
Seattle, WA, 98195} \altaffiltext{2}{Osservatorio Astronomico di
Brera, Milan, Italy} \altaffiltext{3}{Institute of Theoretical
Physics, Univ.of Zurich, CH-8057 Zurich, Switzerland}
\altaffiltext{4}{Dept. of Phys. and Astronomy, McMaster
University, Hamilton, Ontario, L88 4M1, Canada}
\altaffiltext{5}{Department of Phys. \& Astronomy, University of
Pittsburgh, 100 Allen Hall, 3941 O'Hara St, Pittsburgh, PA 1526}
\altaffiltext{6}{Center for Cosmology and Particle Physics, NYU, Meyer
Hall 4 Washington Place 5th Floor New York, NY 10003}
\altaffiltext{7}{Univ. of Victoria, Dept. of Physics and Astronomy,3800 Finnerty Road, Elliot Building, Victoria, BC V8W 3PG Canada}
\altaffiltext{8}{Department of Physics Washington State University
Pullman, WA 99164-2814}
\altaffiltext{9}{Brooks fellow, fabio@astro.washington.edu}

\begin{abstract}

We simulate the formation of a realistic disk galaxy within the
hierarchical scenario of structure formation and study its internal
properties to the present epoch. We use a set of smoothed particle
hydrodynamic (SPH) simulations with a high dynamical range and force
resolution that include cooling, star formation, supernovae (SN)
feedback and a redshift dependent UV background. We compare results
from a $\Lambda$CDM simulation with a $\Lambda$WDM (2keV) simulation
that forms significantly less small scale structure.  We show how high
mass and force resolution in both the gas and dark matter components
play an important role in solving the angular momentum catastrophe
claimed from previous simulations of galaxy formation within the
hierarchical framework.  Hence, a large disk forms without the need of
strong energy injection, the z = 0 galaxies lie close to the I--band
Tully--Fisher relation, and the stellar material in the disk component
has a final specific angular momentum equal to 40\% and 90\% of that
of the dark halo in the $\Lambda$CDM and $\Lambda$WDM models
respectively.  If rescaled to the Milky Way, the $\Lambda$CDM galaxy
has an overabundance of satellites, with a total mass in the stellar
halo 40\% that in the bulge$+$disk system. The $\Lambda$WDM galaxy has
a drastically reduced satellite population and a negligible stellar
spheroidal component. Encounters with satellites play only a minor
role in disturbing the disk.  Satellites possess a variety of star
formation histories linked to mergers and pericentric passages along
their orbit around the primary galaxy. In both cosmologies, the
galactic halo retains most of the baryons accreted and builds up a hot
gas phase with a substantial X--ray emission.  Therefore, while we
have been successful in creating a realistic stellar disk in a massive
galaxy within the $\Lambda$CDM scenario, energy injection emerges as
necessary ingredient to reduce the baryon fraction in galactic halos,
independent of the cosmology adopted.

\end{abstract}


\section{Introduction}
\label{sec:intro}

Within the current paradigm of galaxy formation, realistic disks form
only if gas retains most of its angular momentum gained by torques
from nearby structures while it cools at the center of cold dark
matter (CDM) halos (White \& Rees 1978, Fall \& Efstathiou 1980, Fall
1983, Mo, Mao \& White, 1998).  First simulations of galaxy formation
that included star formation (Lake \& Carlberg 1988, Katz 1992)
provided strong evidence that hierarchical models do create
rotationally supported stellar systems.  However, simulations of
galaxy formation in a full cosmological context have not yet been able
to form realistic disk galaxies: dynamical friction suffered by dense
gaseous lumps and subsequent catastrophic angular momentum loss caused
typical disk scale lengths to come short of those observed (Navarro \&
White 1994).  In addition, high resolution simulations of dark matter
halos in CDM models have far more substructure ("satellites") than
observations; encounters with these satellites could destroy the
stellar disk component of spiral galaxies (Moore et al. 1999).

 If these dark subhalos were the main problem, any mechanism capable
 of reducing the lumpiness of galaxy assembly would lead to realistic
 disks. However resolution simulations (N $\sim 10^4$) also suffer
 from excessive artificial viscosity that would overestimate the
 angular momentum transfer, as was early recognized in previous works
 (e.g (Sommer--Larsen, Gelato \& Vedel 1999).  Poorly resolved disk
 will also suffer from two--body effects which will lead to
 substantial disk angular momentum loss even after its eventual
 formation (Mayer et al. 2001a).

Together, the failure to build galactic disks similar to those
observed and the overabundance of substructure posed a formidable
challenge to the $\Lambda$CDM paradigm, suggesting that a fundamental
ingredient is missing from our understanding of the formation of
galactic and subgalactic structures. As possible solutions, strong SN
feedback (e.g Thacker \& Couchman 2000 (TC) and 2001 (TC01)), an
external UV background produced by QSOs and massive stars (Quinn,
Katz, \& Efstathiou 1996, Benson et al. 2002, Somerville 2002) and
alternatives to $\Lambda$CDM (Spergel \& Steinhardt 2000, among
others) have been advocated.  Among alternative dark matter models,
Warm Dark Matter (WDM) (Pagels \& Primack 1982) emerged as a good
candidate as its lack of power at small scales could prove beneficial
in reducing the number of subhalos and allowing a smoother accretion
of gas, forming larger disks (Bode, Ostriker \& Turok 2001, Col{\'
i}n, Avila-Reese,
\& Valenzuela 2000). Some of these solutions have been explored in
recent studies (e.g. TC01, Sommer--Larsen \& Dolgov 2001,
Sommer--Larsen, Gotz \& Portinari 2003, Abadi et al 2003). 

These most recent works have been successful in forming a rotationally
supported stellar component although in many cases a massive stellar
spheroid is formed as well. In runs that use less than a few thousand
particles to represent the virialized part of a galaxy dark matter
halo (e.g Eke, Navarro \& Steinmetz 2001) the rotating stellar
component does not allow to distinguish a disk.  In TC01, where $\sim$
1.5 10$^4$ dark matter (DM) particles with the halo virial radius
where used, the ``bulge'' component is still almost twice as massive as
the thin disk.  The bulge component decreases to only 50\% of the
stellar disk in the simulation presented by Abadi et al (2003) that
had 3.6 10$^4$ DM particles within the virial radius of their
simulated galaxy.  While these results are encouraging, previous
simulations do not clarify if stronger feedback was necessary, or just
sufficient to avoid systematic loss of angular momentum during the
formation of galactic disks. Runs with larger particle numbers also
had larger disks, suggesting that resolution effects still play a
major role.  Important differences in the results might also arise
from the different merging histories of each individual halo
simulated, the spin of their halos (e.g TC01 simulated a halo with a
larger than average spin which in turn makes it easier to form larger
disks) and the different SN feedback recipes used in various works.

We have performed a new set of
smoothed particle hydrodynamic (SPH) simulations that use a high
number of particles and high force resolution to follow the formation
and internal structure of the stellar component of galaxies.  These
simulations implement cooling, star formation, supernova feedback and
a UV background.  In this paper, we describe the formation and
evolution of a realistic disk dominated galaxies formed in a
cosmological numerical simulation and compare a high-resolution spiral
galaxy formed in 2 different ($\Lambda$CDM and $\Lambda$WDM) flat
cosmologies. Rather than simulate a large sample of galaxies, our
strategy is to achieve the highest possible resolution with the
available computing resources to avoid unwanted numerical
systematics. Our aim is to find a good, clear cut case where a
realistic disk forms in a fully cosmological environment rather than
exploring the full range of galactic morphologies. We also evaluate
the various numerical and physical processes responsible for the
angular momentum loss of the gaseous and stellar components.  We
describe the initial conditions in section 2, we present our results
and discuss their implications in sections 3 and 4 respectively.

\section{Simulation Parameters and Initial Conditions}
\label{sec:sim:cosmo}

\begin{figure}[t]
\centering
\epsfig{file=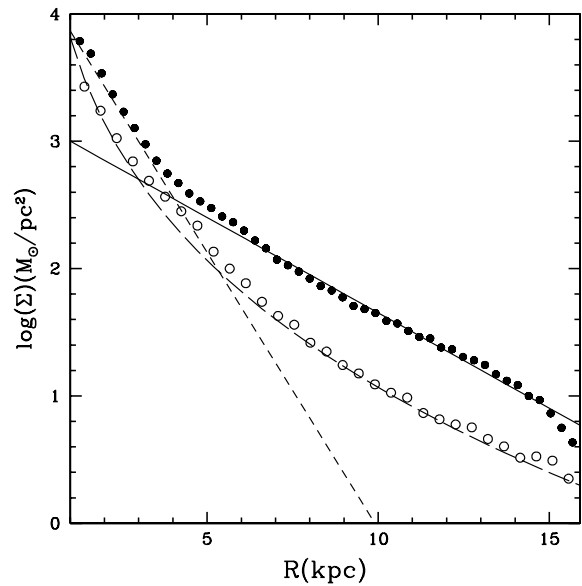,angle=0,totalheight=3.5in,keepaspectratio=true}
\vskip -1mm \figcaption[f1a.eps] { Stellar surface
density profiles for the different stellar components at $z=0$ for the
$\Lambda$CDM galaxy. Filled dots are for the total stellar surface
density profile, open dots for the spheroid only (stars older than 10
Gyr). The black lines are exponential fit to the disk component, with
scale length 3 kpc (3.6 kpc for $\Lambda$WDM). The short dashed lines
are exponential fits to the central stellar mass distribution, with
scale lengths of 1 kpc for both galaxies (the bar and the bulge are
both contributing), and the long dashed line is a de Vaucouleurs fit
to the spheroid, with scale lengths of 800 pc ( $\Lambda$CDM galaxy)
and 500 pc ($\Lambda$WDM galaxy)}
\vskip -3mm
\end{figure}

\begin{figure}[t]
\centering
\epsfig{file=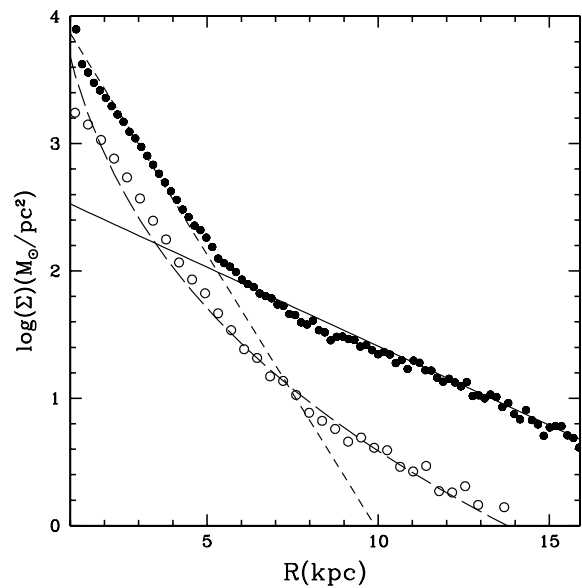,angle=0,totalheight=3.5in,keepaspectratio=true}
\vskip -1mm \figcaption[f1b.eps]{Stellar surface
density profiles for the different stellar components at $z=0$ for the
$\Lambda$WDM galaxy. Filled dots are for the total stellar surface
density profile, open dots for the spheroid only (stars older than 10
Gyr). The black lines are exponential fit to the disk component, with
scale length  3.6 kpc. The short dashed lines are exponential fits
to the central stellar mass distribution, with scale lengths of 1 kpc
for both galaxies (the bar and the bulge are both contributing), and
the long dashed line is a de Vaucouleurs fit to the spheroid, with
scale lengths of 500 pc}.
\vskip -3mm
\end{figure}

We adopted a flat $\Lambda$-dominated cosmology: $\Omega_0=0.3$,
$\Lambda$=0.7, $h=0.7$, $\sigma_8=1$, shape parameter $\Gamma=0.21$,
and $\Omega_{b}=0.039$ (Perlmutter et al. 1997, Efstathiou
et al. 2002).  Power spectra were calculated using 
 the CMBFAST code to generate transfer functions (Zaldarriaga
\& Seljak 2000). The $\Lambda$WDM model can be characterized in terms
of the reduction in fluctuations on dwarf galaxy scales.  For our
choice of 2 keV neutrinos the fluctuations on scales smaller than
$\sim 10^{9.5} M_{\odot}$ are vastly reduced.  We retained the phases
of the waves that describe the initial density field. These allowed us
to discriminate between the different effects of the two power spectra
used.

From a 100 Mpc box, low resolution simulation, we selected and
resimulated (Katz \& White 1993) at higher resolution a ``typical''
spiral galaxy candidate halo with mass 3 $\times 10^{12}M_{\odot}$ and
a quiet merger history after $z\sim 2.5$. The halo is relatively
isolated and its formation time (defined as the main progenitor
achieving 50$\%$ of its final mass) is z = 0.75. Its spin parameter,
$\lambda$ is 0.035 (0.03 for $\Lambda$ WDM) close to the average value
for cosmic halos (Gardner 2001). Note that due to the finite size of
the box the actual $\sigma_8$ in the simulation volume is actually
$\sim$ 0.9 at the present time.  The above constraints are consistent
with the Milky Way forming several Gyrs ago (Wyse 2002) and
predictions from semianalytical models (Baugh, Cole \& Frenk 1996). A
large starting box is important as lack of large scale power would
significantly reduce the amount of torques on collapsing halos.  To
run the simulations, we used a parallel treecode+SPH with multiple
timesteps: GASOLINE described in detail in Wadsley, Stadel \& Quinn
(2003) and Stadel, Wadsley \& Richardson (2001).  The code treats
artificial viscosity as suggested in Balsara (1997). The energy
equation is computed asymmetrically (Springel \& Hernquist 2002,
Evrard 1988, Monaghan 1992 ).  This approach avoids the energy
conservation and negative energy problems of the arithmetic and
geometric forms  and converges to the high resolution answer faster
than other proposed methods (Benz 1990, Springel and Hernquist 2002).

The high resolution region has dark matter and gas particles of mass
2.32$\times$10 $^{7}$ and 3.44$\times$10 $^{6}$ M$_{\odot}$
respectively. The rest of the simulation box was re-sampled at
increasing particle masses for a total of nearly 1.1 million dark
matter particles and 2.2$\times$10$^5$ gas particles.  For all
particles species in the high resolution region, the gravitational
spline softening, $\epsilon(z)$, evolved comovingly starting at z=100
until z=9, and remained fixed at 1 kpc from z=9 to the present.  This
choice of $\epsilon(z)$ reduces two body relaxation at high z in small
halos. The value chosen is a good compromise between reducing two body
relaxation and ensuring that disk scale lengths and the central part
of dark matter halos will be spatially resolved (see Diemand et
al. 2002 for a number of relevant tests).  Moreover, a constant
softening in physical coordinates is more appropriate at low redshift
when stellar structures detached from the Hubble flow.
Integration parameter values were chosen as suggested in Moore et
al. (1998) and then confirmed in Power et al. (2002).  Most of the
stars formed in the simulation required 32,768 integration steps to
$z=0$, with a significant fraction of them requiring four times as
many steps. At comparable or better force resolution, particle number
within the virial radius is three times as in Navarro \& Steinmetz
2002, eight times as in TC01, three times as in Abadi et al. 2002 and
comparable to the highest resolution run reported in Sommer--Larsen,
Gotz \& Portinari 2003).
 
Two main SPH simulations were run, one for each cosmology. Both runs
included: Compton and radiative cooling, assuming a gas of primordial
composition; star formation and SN (type I\&II) feedback, treated
following the prescription described by Katz (1992), where stars spawn
from cold,  Jeans unstable gas particles in regions of
converging flows. The star formation efficiency parameter $\epsilon$ was set
to 0.15, but with the adopted scheme the exact value of  has only a minor
effect on the star formation rate (Katz 1992). SN enrich the
surrounding gas according to the respective metal yields. Star
particles inherit the metal abundance of the parent gas
particles. However we did not allow for diffusion of metals between
gas particles.

After a gas particle is less than 10\% of its initial mass due to
multiple star formation events, it is removed and its mass is
re-allocated among its gas neighbors. Up to six star particles
particles are then spawn for each gas particle in the disk.  The
adopted ``minimal'' feedback recipe dumps energy from SN into thermal
energy of the nearest 32 gas neighbors. As energy is quickly radiated
away in dense gaseous regions this feedback recipe has a weak effect
on star formation (TC). We explicitly choose a fairly massive galaxy
and of a weak feedback to explore a regime where feedback is likely to
play a minor role.

The evolution and strength of the uniform UV background from QSOs
followed Haardt and Madau (1996) and Haardt (2002, private
communication).  To complement our study, we performed dark matter
only simulations of this same halo at similar and eight times higher
resolution and 50\% better spatial resolution.  These simulations
showed the main properties of the galaxy's dark matter halo (inner
profile, spin) are not changed substantially by the introduction of
WDM, as the galaxy mass is much greater than the free--streaming mass
at 10$^{9.5}$M$_{\odot}$.

\section{Results}

 Within the $\Lambda$CDM galaxy virial radius (defined by $\delta
 \rho/ \rho_{crit}$ = 97.1 at $z = 0$) there are $\sim 1.2 \times 10^5$
 dark matter, $5 \times 10^4$ gas and $6 \times 10^5$ star
 particles. The main parameters of the galaxies at $z=0$ are described
 in Table 1.  Star formation (Fig.5) begins at $z=10$, before
 the external UV background kicks in at $z \sim 7$.  The star formation
 rate (SFR) peaks around $z=5$ at  80$\Msun$/yr while for the
 $\Lambda$WDM run the peak is delayed and is only 45 $\Msun$/yr at $z=4$
 due to the  effect of removal of power at small scales.

In both cosmologies the bulge of the galaxy forms through a series of
 rapid major merger events that end by $z=2.7$, turning mostly low
 angular momentum gas into stars. At $z = 2.5$ the hot gas (T$>10^5$K)
 left within 50 kpc has a higher specific angular momentum than the
 dark matter (van den Bosch et al. 2002).  Soon after the
 last major mergers, the disk starts forming from the infall of this
 high angular momentum gas.  At this stage the disk of the WDM galaxy
 is thinner than that formed in the $\Lambda$CDM one. In both
 simulations the final 3-4 Gyrs of evolution are marked by a bar
 instability, seen also in our own Milky Way (Cole \& Weinberg 2002),
 and repeated encounters of the galaxy with a few relatively massive
 infalling satellites.

\subsection{The Galaxy Bulge and Disk components}

The resulting galaxies show that we successfully formed extended
and fairly old stellar disks in $\Lambda$-dominated hierarchical
models.  In Fig.1 and Fig.2 the stellar spheroid (including bulge and halo
stars) is defined as the sum of all stars born before or at the time
of the last major merger; this selects low angular momentum stars for
which $v_{rot}/\sigma < 0.5$ (where $v_{rot}$ is the tangential
velocity and $\sigma$ is the radial velocity dispersion), with the
stars in the central few kpc having a $v_{rot}/\sigma$ as low as
0.1-0.2, similar to what inferred for the stars belonging to the
spheroid of the Milky Way (Minniti 1996).  We were not able to use a
kinematic criterion in order to separate the central ``bulge'' from
the diffuse extended spheroid --- this resembles what found for our own
Milky Way, where only a combination of metallicities and kinematics
allows a distinction (Minniti 1996).  We fit a de Vaucouleurs profile,
characterized by a scale length $R_b$, to the old stellar
distribution, and then we define the mass of the bulge as twice the
mass contained within $R_b$ ($R_b$ by definition corresponds to the
half-mass radius of the system if we rely on the fit). This yields the
total bulge masses indicated in Table 1 (precisely the ratio between
the mass of the bulge and the mass of the disk, $B/D$ is indicated
there), and the bulge scale length is about 800 pc and 500 pc in,
respectively, the $\Lambda$CDM and the $\Lambda$WDM run.

At $z=0$ a flat, rotationally supported stellar component with
$v_{rot}/\sigma > 1$ extends out to almost $20$ kpc (Fig.1 and Fig.2).  This
disk component is thinner and dynamically colder (lower
$v_{rot}/\sigma$) in the $\Lambda$WDM model. The decomposition of the
stellar distribution into a dynamically separate disk and a bulge is
complicated by the presence of a third component, namely a prominent
bar-like structure.  The latter is significantly more elongated and
easier to distinguish in the $\Lambda$WDM galaxy.  In the $\Lambda$CDM
galaxy a "fat" bar or oval distortion typical of early-type galaxies
(Athanassoula 2003) is first seen at $z=0.6$ and then is progressively
thickened by tidal heating from infalling satellites as well as by
numerical heating, becoming progressively harder to distinguish.
However, its signature is clearly evident in the kinematics; if we
select all stars younger than 10 Gyr, thereby born after the last
major merger, the kinematics are disk-like from 20 kpc going toward
the center ($v_{rot}/\sigma > 1.5$, but become hotter ($v_{rot}/\sigma
\simeq$ 1) at 3-4 kpc from the center, the transition being sharper in
the case of the $\Lambda$WDM galaxy (Fig.2).

As mentioned above, the stellar density profiles of both galaxies at
$z=0$ is thus the sum of the old spheroid formed by the last major
merger and the much younger stellar bar in the inner few kpcs, whereas
the cold disk dominates in the outer part (Fig.1 and Fig.2). A double
exponential provides a very good fit to the data. However, as from
Fig.1 it is worthwhile noting that a straightforward B/D decomposition
based on the inner and outer component resulting from such fit would
give erroneously large bulge masses and sizes, counting the thin,
rotating and relatively young stars in the bar as part of the
bulge.The overestimate of the bulge mass would be particularly severe
in the $\Lambda$WDM galaxy.

Given the initial halo spin ($\lambda$ = 0.03-0.035) the disks have
scale lengths comparable to those predicted by theoretical models in
which baryons conserve their specific angular momentum during infall
(see Mo, Mao and White 1998, which keeps into account the adiabatic
contraction of the dark matter halo due to the baryonic build up).
The average age at the solar radius is 9-10 Gyrs for the $\Lambda$CDM
run (8 for the $\Lambda$WDM run), in agreement with current
observational estimates (e.g. van den Bergh 1996) In Fig.1 we also
plot the surface density profile of the old spheroid component -
clearly this accounts only in part for the inner steepening. The bar
contributes for the remaining part; the connection between the
presence of a bar and a steep inner stellar profile has been widely
shown by a number of numerical simulations on bar formation and
evolution (Sparke \& Sellwood 1987; Pfenniger \& Friedli 1991, Mayer
\& Wadsley 2003).  Interestingly, even in the case of the Milky Way
the debate is still on-going on whether the central, kinematically hot
component usually identified as the bulge is in part or even entirely
a stellar bar (Zhao, Rich \& Biello 1996).

The final spiral galaxies are more massive than M31 and the Milky
Way. In both cosmologies the circular velocity (defined as v$_c$ =
(M(r$<$R)/R)$^{1/2}$) at the virial radius is $\sim$ 185 km
sec$^{-1}$, while recent estimates (Klypin, Zhao \& Somerville 2002)
yield a v$_c$ at the virial radius around 150-160 km sec$^{-1}$ for
our MW. At at 2.2 disk scale lengths (a distance typically used for
observational samples) v$_{2.2} = 320$ km sec$^{-1}$ (280 km
sec$^{-1}$ for $\Lambda$WDM, (see Fig.3 and Fig.4). As in other works,
some of them including strong SN feedback (Abadi et al. 2002) the
rotation velocity peaks even higher and in regions very close to the
galaxy center that are only poorly resolved. These unrealistic high
central velocities, linked to a very concentrated spheroidal
component, are present in runs with and without SN feedback. They are
a problem of most current galaxy formation runs.

\begin{figure}[t]
\centering
\epsfig{file=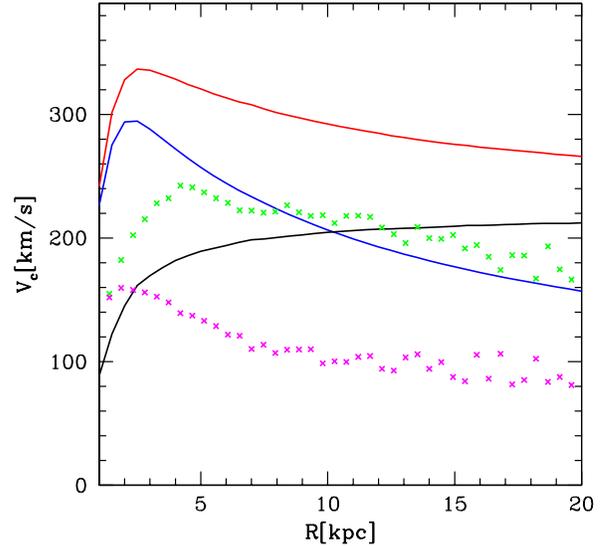,angle=0,totalheight=3.6in,keepaspectratio=true}
\vskip -1mm \figcaption[f2a.eps] {Rotation curve (defined as
$V_c = \sqrt{M((r<R)/R}$) for all components (continuous red line), star
(blue) and dark matter (black) for the $\Lambda$CDM galaxy at
z=0. The stellar kinematics are also shown; the rotational velocity is
indicated by the green crosses, while the (azimuthally averaged) radial
velocity dispersion in cylindrical coordinates (the z-axis being
chosen along the angular momentum vector of the disk) is magenta}
\vskip -5mm
\end{figure}

\begin{figure}[t]
\centering
\epsfig{file=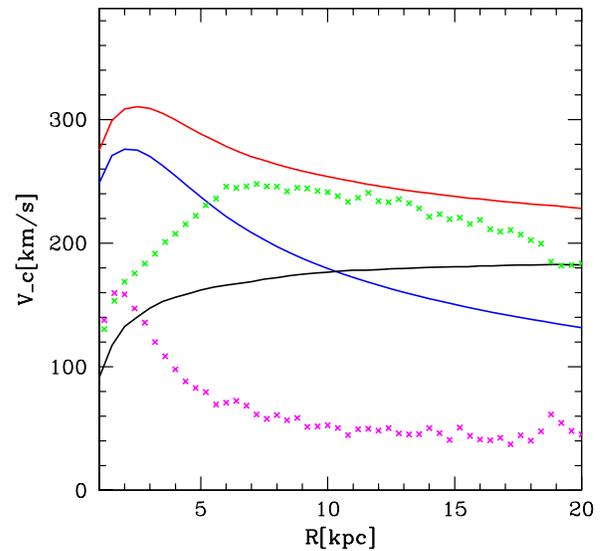,angle=0,totalheight=3.6in,keepaspectratio=true}
\vskip -1mm \figcaption[f2a.eps] {Same as for Fig.3, but for $\Lambda$WDM}.
\vskip -5mm
\end{figure}

By the present time, star formation has turned about 64\% (55\% for
$\Lambda$WDM) of gas inside the virial radius into stars, with
$\Lambda$CDM having a slightly larger bulge component.  At the final
time only a small amount of gas within the virial radius remains in
the cold phase (1.5\% compared to 2\% at z=0.25 and 9\% at z=1 in the
$\Lambda$CDM). Most of it is in the central disk. The $\Lambda$WDM has
three times as much cold gas in the disk.  The exponential fits to the
stellar disk mass distribution yield scale lengths of $\sim 3$ kpc for
the $\Lambda$CDM and $\sim 3.6$ kpc for the $\Lambda$WDM. Bulge scale
lengths are considerably smaller (Fig.1). 

The latter fits to
the stellar distribution are obtained excluding the inner,
bar-dominated region from the disk.  Strictly speaking the bar is part
of the disk since it was formed from disk material that underwent
non-axisymmetric instabilities. Therefore we also tried fitting a
single exponential over the whole stellar component with age $< 10$
Gyr; given the remarkable change of shape of the stellar density
profile at a few kpcs from the center a single exponential provides a
poor description. Nevertheless, it is worth noticing that, by doing
this, the scale lengths drop to 2 kpc and 1.5 kpc for, respectively,
the $\Lambda$CDM and the $\Lambda$WDM galaxy (as expected the change
is more marked for the $\Lambda$WDM galaxy, which exhibits a stronger
and longer bar).  In the existing literature (see Carollo et
al. 2001, Debattista,Carollo, Mayer \& Moore 2003) the  flatter component
of the stellar profile is always included in the definition of the
disk, while bars are often identified with a central bulge. These
results support our double fits and the resulting scale lengths.

By measuring the ratio of
the specific angular momentum inside the virial radius of the halo in
a dark matter only version of the simulations to that of the stellar
disk at $z=0$, we found that $\sim 60 \%$ of the specific angular
momentum has been lost in the $\Lambda$CDM galaxy and only $10\%$ in
the $\Lambda$WDM galaxy. The expected disk scale-length, $R_d$,
based on the models by Mo, Mao \& White (1998) (see also Fall (1983)
and Navarro \& Steinmetz (2000)) is 3.4 kpc ($\Lambda$CDM) assuming a
50\% loss of the specific angular momentum, thus very close to what we
find, once the halo's spin and the adiabatic contraction of the dark
halo due to the formation of both the bulge and the disk are taken
into account. By neglecting the latter a much larger disk scale
length, around 5 kpc, would be predicted by the analytical approach.
A comparison between analytical predictions by Mo, Mao \& White (1998)
and observed disk galaxies (Courteau 1997) shows that the scatter in
disk scale length measured by Courteau (1997) is more than a factor of
3 across the entire circular velocity range.  A galaxy with a scale
length of about 3 kpc and v$_{2.2}$ 320 km/s stays lower than the mean
on the relation predicted by the analytical model of Mo, Mao \& White
(1998), but still quite consistent with such data sample which
consists mostly of Sb and Sc. However, to match the average properties
of disk galaxies of type Sb and later a smaller angular momentum loss,
or a higher initial halo angular momentum will be required. Indeed the
$\Lambda$CDM galaxy would most likely be classified as Sa or Sab due
to its prominent bulge.

Using age and metallicity information from every star particle created
and Single Stellar Population models (Tantalo, Chiosi \& Bressan 1998)
we find that at $z=0$ the $\Lambda$CDM galaxy has $M_I = -24.3$ (including disk
and bulge) and lies close to  the Tully--Fisher relation (Giovanelli et al.
1997), Its disk B--R color is 1.05, while the average B--R
in Courteau (1997) sample is 0.8.

\begin{figure}[t]
\centering
\epsfig{file=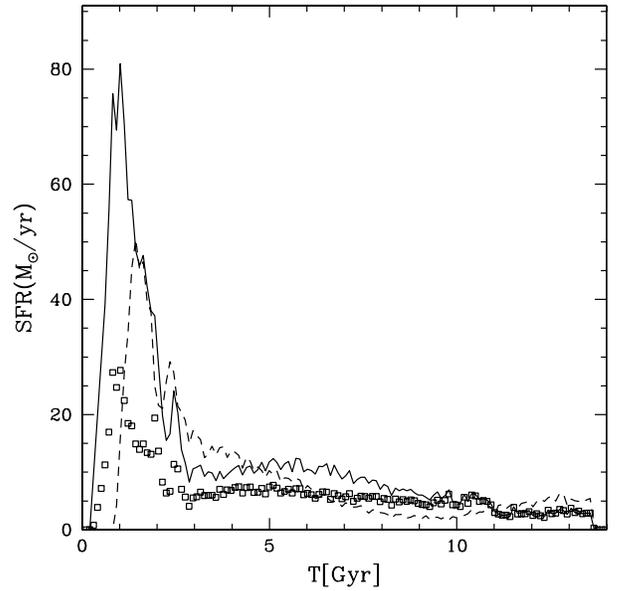,angle=0,totalheight=3.6in,keepaspectratio=true}
\vskip -1mm \figcaption[f3.eps] {SFR of the $\Lambda$CDM
(continuous) and $\Lambda$WDM (dashed) galaxies. The line with small
circles shows the SFR of the $\Lambda$CDM disk.}
\vskip -3mm
\end{figure}

Internal instabilities of the stellar disk and perturbations by
satellites cause structural and kinematical changes in the disk,
heating it substantially after most of its mass has been assembled by
z$>0.7-0.5$. As mentioned above, the inner, colder regions where Q
(Toomre's instability parameter) is initially $< 2$, become unstable
to a bar-like mode at $z \sim 0.6$; part of the gas loses angular
momentum to the stellar bar, flows to the center and becomes the site
of a final episode of star formation.  The disk of the $\Lambda$WDM
galaxy stays thinner because tidal heating by satellites is largely
suppressed.  We tested (section 4) that the $\Lambda$CDM galaxy
becomes bar-unstable mostly because of its fairly massive disk
(Efstathiou, Lake \& Negroponte 1982, Athanassoula \& Sellwood 1986).

\begin{figure}[t]
\centering
\epsfig{file=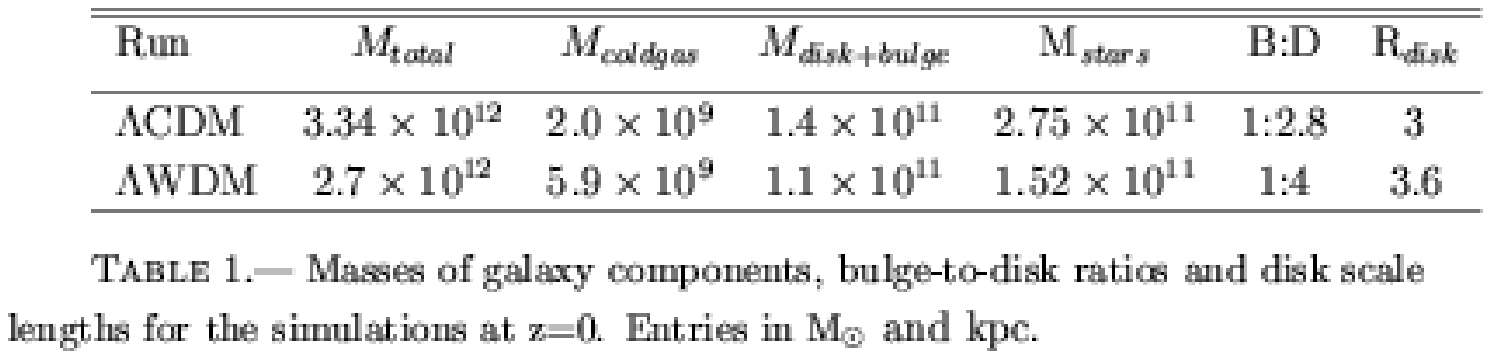,angle=0,totalheight=1.in,keepaspectratio=true}
\vskip -3mm
\end{figure}

\subsection{The Galaxy Halo}

As expected in $\Lambda$CDM models, a rich amount of substructure
develops and forms a substantial stellar component due to a
combination of weak feedback and early star formation. In excess of
what is observed for Andromeda or the Milky Way (Mateo 1998) the
$\Lambda$CDM model has at least thirty satellites with stellar masses
larger than 1.35 $\times$10$^9 M_{\odot}$. Consistent with its lack of
small scale power $\Lambda$WDM galaxy only has seven.  20\% of the
stars formed within the virial radius of the $\Lambda$CDM galaxy are
in the satellites, and another 15\% are in the halo within r$>$20
kpc. In comparison, due to its lack of power at small scales only 15\%
of stars are in the $\Lambda$WDM satellite system (3 of them are
within 35 kpc of its center) and only 0.2\% of the stars are in the
outer part of the halo.  The most massive (and best resolved)
satellites show a large variety of star formation histories (Fig.6),
in both cosmologies.  Their star formation ceases a couple of Gyrs
after entering the halo of the primary galaxy.  At $z=0$ the stellar
component of the dwarfs that entered the galaxy halo at $z \ge 0.5$
are supported by velocity dispersions, while those outside or that
have just fallen in have recognizable disks, as expected if dynamical
encounters are the major drivers of satellites morphologies as in the
``tidal stirring'' scenario (Mayer et al. 2001).  In both models, the
diffuse halos contain the oldest stars, but they also grow
continuously from material stripped after the last major merger that
gives rise to the bulge and from the debris of disrupted satellite
until late epochs.

\begin{figure}[t]
\centering
\epsfig{file=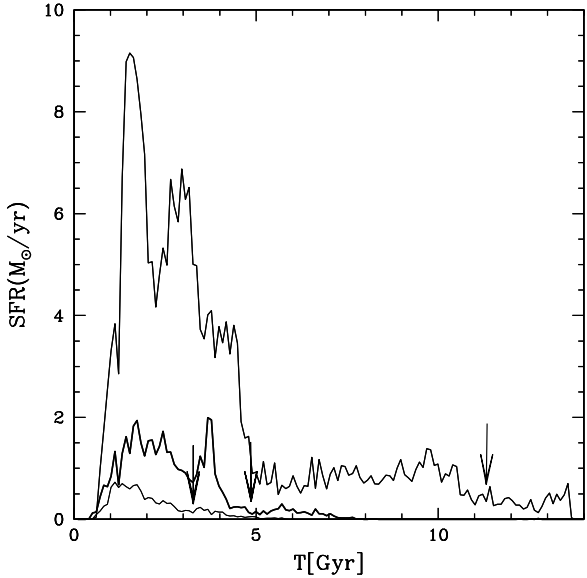,angle=0,totalheight=3.5in,keepaspectratio=true}
\vskip -1mm \figcaption[f4.eps] {Star
Formation histories for three satellites of the $\Lambda$CDM galaxy at
z$=$0. Arrows mark the time they entered within R$_{vir}$. Satellites'
SFRs peak in correspondence of merger events and pericentric
passages}.
\vskip -3mm
\end{figure}

The amount of baryons (including stars, cold and hot gas) inside
the virial radius is within a few percent of the original cosmic
abundance. The X-ray flux at z=0 is $\sim$ 0.5 10$^{42}$ erg/sec in
both cosmologies, this value doubles if a metal abundance of 0.3 solar
is assumed for the gas in the halo.  This is more than expected from
observations, for example the large disk galaxy NGC2841 has a
bolometric emission of only 1.3 10$^{41}$ (Benson et al. 2000, Murali
2000). As expected with the weak feedback adopted and the long
cooling times in the outer part of this relatively massive halo a hot
gas phase is always present within the virial radius, and X-ray
luminosity evolution is weak to z=1. The overabundance of halo baryons
(in all its various subcomponents, hot gas, satellites and diffuse
stellar halo) is clearly a problem of both cosmologies.  While the WDM
galaxy has a substantially smaller stellar halo component, possibly
in agreement with observations, its X-ray emission is high and very
similar to the$\Lambda$CDM one. This is expected as reduced power at
subgalactic scales and does not affect the total mass and the
accretion rate of halo baryons. According to the existing literature
and the work presented here, energy feedback, reduced amount of small
scale power and increased resolution tend to increase the size of
galactic disks, but the amount of hot gas in halos of large galaxies
is likely only sensitive to feedback. X--ray observations sensitive to
sub keV temperatures of nearby massive galaxies sensitive would set
powerful constraints on the amount of energy injected by different
processes into the baryonic component during the formation of the
galaxy, as they currently do on the larger scale of small groups
(Borgani et al. 2002), pointing to at least 0.5keV/baryon. As an
alternative  Toft et al. (2002) suggest that the excess X-ray
emission could be reduced allowing a detailed treatment of cooling by
metal lines that would reduce the amount of hot gas left in the
halo. Current global estimates of galaxy formation
efficiency set it at about 10\% (Bell et al. 2003) which is 
significantly lower than the high star formation efficiency in our
runs, suggesting that  ejection of baryons from the halos of disk
galaxies is likely necessary to avoid excess X-ray emission from
galaxy halos. The hot gas component of the Milky Way has just recently
been detected (see Sembach 2003) and future observations will provide
useful constraints on its properties.

\section{Robustness of Numerical Results: solving the angular momentum problem}

The main finding of this work is that the angular momentum and
consequently the scale length of the stellar disk formed in our
simulations is comparable to that of real galaxies of similar
luminosity.  This result was not obtained adopting the usually
advocated strong feedback to decrease the lumpiness of the gas being
accreted, but rather using good mass and force resolution. Is then the
addition of feedback still a necessary ingredient to obtain realistic
galaxies? How much is the formation and evolution of disk stellar
systems affected by limited resolution?

It is now widely recognized that 10$^{5-6}$ particles with a force
resolution close or better than 1\% R$_{vir}$ are necessary to
reliably simulate the formation and the internal structure of a single
dark matter halo (Ghigna et al 2000, Fukushige \& Makino). It is then
unlikely that the much more complex formation of disk galaxies could
be reliably simulated with significantly less particles.  A comparison
with our higher resolution dark matter only runs shows that the mass
function of dark matter subhalos becomes severely incomplete below
30--35 km sec$^{-1}$. On the positive side this implies that at our
resolution we are resolving the progenitors mass range that
contributes most of the mass in the accretion history of the central
galaxy and most of the mass in the subhalo population.  With more than
10$^5$ particles within the virial radius, our simulations accurately
follow the evolution of the hot gaseous phase (see tests in Frenk
et al. 1999, Borgani et al. 2002, Wadsley, Stadel \& Quinn 2003).

The final angular momentum of a stellar galactic disk is set
first by the details of hierarchical accretion onto the parent halo
and the subsequent cooling of high angular momentum gas.  Once a disk
of baryonic material is formed, disk instabilities might redistribute
the original angular momentum of the disk both internally and to the
dark matter halo (Weinberg \& Katz, Valenzuela \& Klypin 2003).  Both
stages are potentially affected by numerical viscosity (present in the
SPH approach to describe shocks) and numerical relaxation introduced
by noise in the global potential due to small particle numbers.  These
effects reduce the angular momentum of the disk by artificial
dissipation or by transfer to the halo component respectively.

To measure the importance of the numerical effects mentioned in the
above paragraph, but independently of the hierarchical build up, we
built an isolated galaxy system (following Springel \& White 1999)
with the same structural parameters (mass, profile, fraction of mass
in the disk and in the bulge, relevant scale lengths, fraction of cold
gas in the disk) as our $\Lambda$CDM galaxy at z=0.6, i.e.  soon after
the formation of the disk. Results can likely be extended to the
$\Lambda$WDM galaxy.  All components were made of ``live'' particles
This galaxy was evolved in isolation for 6 Gyrs.
We run different version of it changing the number and
accordingly the mass and softening of the stellar, dark matter and gas
particles (see Fig.7). Softening for the different component in the
different realizations was scaled as $\epsilon_{test}$=
$\epsilon_{cosmo}$ (m$_{test}$/m$_{cosmo}$)$^{1/3}$, were
$\epsilon_{cosmo}$ and m$_{cosmo}$ were the softening and masses used
in the main cosmological runs for each component.

\begin{figure}[t]
\centering
\epsfig{file=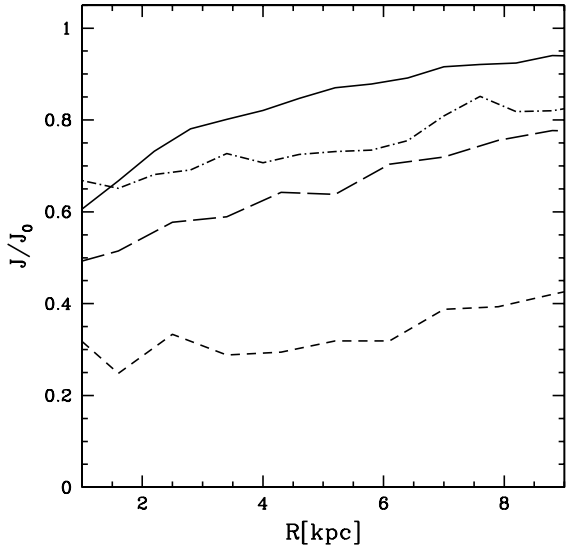,angle=0,totalheight=3.6in,keepaspectratio=true}
\vskip -1mm \figcaption[f5.eps] {Angular momentum loss in an isolated
disk galaxy model with structural parameters as our $\Lambda$CDM run
at z = 0.6. This model was run for 6 Gyrs ( equivalent to the present
time).  The y axis shows the fractional angular momentum loss for all
the baryonic material in the disk as a function of radius.  Continuous
line:N$_{DM}$ = 100000, Nstar=200000, Ngas=5000 (same as in the cosmo
run).  Our choice of spawning a large number of star particles
(typically 6 per gas particle) in the cosmological run reduces
internal heating of the stellar disk.  dotted short dashed: N$_{DM}$
and Nstar reduced by a factor of five. Long dashed: Nstars reduced by
a factor of 25. short dashed:N$_{DM}$= 4000, Nstar=8000, Ngas=1000. At
the lowest resolution the disk undergoes the catastrophic angular
momentum loss reported in early simulations}.
\vskip -3mm
\end{figure}

In our lowest resolution realization (25 times less particles
than in our cosmological runs), the number of halo (DM) particles and
particles representing stars in the disk goes below a few thousands and
the disk (defined as all the particles originally assigned to the disk
component) loses almost 70\% of its initial angular momentum over 8
Gyrs.  A disk structure is not visible at the final (present) time
(Fig.7 short dashed).  Angular momentum loss is greatly reduced as the
number of DM particles first and then number of particles in the
stellar disk is increased (long dashed and dot dashed).  At the
resolution of our cosmological runs (continuous line) the loss is down
to only 10\% across most of the disk, reaching 30\% in the inner 2
kpc. Even in this isolated model a bar forms in the disk. Finally, we
verified that when the collisionless components are well resolved then
gas angular momentum loss due to artificial viscosity becomes less
than 10\% over several Gyrs if the disk gas component is represented
by at least a few thousand particles (in agreement with Navarro \&
Steinmetz 2000).  We expect numerical effects will continue to
decrease as the numerical resolution (both spatial and in time) is
increased.

To verify that our findings are applicable in a full cosmological
 context where the hierarchical build is included, we also run a low
 resolution realization (1/8th of the particles and twice as large
 $\epsilon$ of the cosmological $\Lambda$CDM galaxy described in our
 work. In this situation additional resolution problems could be
 present, as the first individual structures that form are necessarily
 made of few particles. Artificial viscosity could show up not only
 in cold gas lumps, but also in poorly resolved shocks and numerical
 effects could cascade down as the galaxy main body is assembled.
 This also implies that, at the same mass and force resolution, the
 $\Lambda$CDM galaxy with its larger substructure mass range, is
 intrinsically harder to simulate than the $\Lambda$WDM one.  As
 expected, due to the low number of particles and the very poor
 spatial resolution by z=0 in this low resolution run, cold gas and
 stars within 20~kpc have lost 90\% of their initial angular momentum
 compared to the dark matter and no strong stellar disk component is
 present, although the central part of the stellar distribution is
 rotationally supported.  This low resolution run is in qualitative
 agreement with Eke, Navarro \& Steinmetz 2000, where no disk like
 morphology was observed in their simulated galaxies. Our results
 imply that the disk galaxy described in TC01 (their run was stopped
 at z=0.5) would likely show a much weaker disk component if evolved
 to the present time.

To show the effects of resolution on a galaxy with different
 mass and merging history from that in our main runs we re--simulated
 to z$=2$ (with similar techniques as described for the main runs) a
 small dwarf galaxy sized halo (9$\times$10$^{10}$M$_{\odot}$) at the
 outskirts of our MW halo at 8 times better mass resolution. As the
 number of DM particles within R$_{vir}$ (gas particles were increased
 accordingly) increased from $\sim$ 3000 to almost 2.5~ 10$^4$ and the
 softening was set at 0.5 kpc, the stellar component of this dwarf
 galaxy became more rotationally supported as v/$\sigma$ changed from
 0.46 to 0.77 and a gaseous disk became apparent. The shape and
 amplitude of the SFR(z) did not change substantially, showing that we
 are qualitatively capturing the events that drive star formation in
 the small galactic satellites of our main runs.

To summarize, at the best resolution for our main runs numerical
effects are likely responsible for the angular momentum loss in the
$\Lambda$WDM disk (about 10\%).  As described in section 3 the
$\Lambda$CDM disk loses significantly more angular momentum than its
equivalent in $\Lambda$WDM.  The only difference in the two runs are
the initial power spectra, with $\Lambda$CDM having significantly more
substructure below 50-100Km/sec. This result suggests that {\it at the
current resolution}, and as suggested by early works the intrinsic
clumpiness of the hierarchical formation process leads to significant
angular momentum loss, likely through interactions between the
infalling lumps (Lake \& Carlberg 1986) before the disk has formed.
However, at our resolution, this is angular momentum loss is far from
catastrophic as the fraction of mass contained in surviving sublumps
within a larger parent halo is only of the order of 15\% (Ghigna et al
1998).  A large fraction of mass is tidally stripped from the subhalos
as soon as they first go past pericenter (Mayer 2001).  This material
typically loses only a small fraction of its initial angular momentum
(Colpi, Mayer \& Governato 1999.)  If this high fraction of stripped
material is also typical of the gas accreted by the disk then most of
it actually forms from diffuse material cooling from the hot
halo, rather than by individual lumps.

Our results suggest that the large angular momentum loss or the
lack of well defined stellar disks reported in early works might have
resulted from poor force or mass resolution. The tests done with the
isolated galaxy model suggest that two-body heating by massive halo
particles would be the main cause of angular momentum loss at low
resolution if small softenings were used.  As done in some previous
works, two-body heating can be suppressed, by using a larger softening
without increasing the number of particles. However, even softenings
as small as those values used here are barely enough to resolve the
disk and avoid spurious results as for its morphological evolution;
suppression of disk self-gravity by a softening larger than the disk
scale length would inhibit processes like bar formation that are
essential ingredients in disk formation and evolution (Romeo 1991,
1994; Weinberg \& Katz 2002; Valenzuela \& Klypin 2003; Mayer \&
Wadsley 2004).

We also evolved another isolated galaxy model with the same structure
and numerical resolution as the $\Lambda$CDM galaxy, but with a
stellar disk only 50\% as massive. This disk does not show the bar
instability present in the other runs, and its angular momentum loss
is slightly lower.  This experiment also implies that disk-satellite
interactions have only a limited effect on the development of the bar
instability and in the angular momentum loss of the disk.

 Finally, it is likely that our adopted feedback recipe is too
 simplistic, and although it models the build up of the disk
 population well, it does not stop star formation at high redshift in
 small halos creating an excess of massive satellites in the
 $\Lambda$CDM galaxy The paucity of cold gas in our final state
 suggests that our star formation algorithm is likely too efficient,
 at least compared to the universal value of 10\% or less (Bell
 et al. 2003) although is difficult to quantify such a statement with
 only one run.

\section{Summary and Discussion}

We simulated the formation of a disk galaxy in a cosmological context
to the present time with a high force and mass resolution.  Without
resorting to strong feedback recipes, we successfully formed an object
with bulge-to-disk mass ratios, scale lengths, current star formation,
disk age and dynamical properties representative of those observed in
present day massive spirals. Our results show that the large angular
momentum loss reported in some early works was in at least in part due
to insufficient mass or force resolution. High resolution plays a
significant role in simulating correctly the formation of disk
galaxies in a cosmological context, independently of the SN feedback
recipe adopted.

We estimate that at a resolution of
$\sim 10^5$ particles within $R_{vir}$ for each of the DM, gas and
stellar components numerical effects still account for 10\% of the
disk angular momentum loss during its formation and evolution.

WDM improves on the standard$\Lambda$CDM model on a few aspects and
leads to testable differences.  Namely it reduces the number of
satellites and the galaxy forms with a smaller bulge and almost no
stellar halo component. Its disk is more gas rich at the
present time and has colder kinematics (lower $v_{rot}/\sigma$).
However, WDM fails to solve the issue for which it was
introduced: WDM does not make substantially more extended disks.

The presence of a substantial fraction of hot gas in the halo,
independent of the cosmology adopted, results in an X--ray emission
higher than estimates for galaxies of similar v$_{c}$ to those in our
study. Numerical and semi-analytical results (e.g Borgani et al. 2002,
Babul et al. 2002) show that significant energy injection leading to
an entropy floor of 50--100 keV/cm$^2$, is needed on the mass scale of
small groups (just a factor of ten higher than the mass of the galaxy
in our study) to reduce the fraction of hot baryons and reconcile
models with the observed X--ray Luminosity--Temperature relation.
However, more numerical studies including cooling and detailed star
formation are needed.

There is little evidence for SN energy injection of this magnitude in
dwarf galaxies (Martin 1999), but it is possible that a combination of
SN and AGN activity linked to bulge formation (Binney, Gerhard \& Silk
2001) could simultaneously reduce the baryon fraction in the halo and
quench early star formation in the disk and the satellites, making a
large fraction of them ``dark''. If energy injection is required
regardless of cosmology, then the need to invoke non standard dark
matter models will be substantially weakened.



\acknowledgments

Simulations have been run at the Arctic Region Supercomputing Center
and the Pittsburgh Supercomputing Center. FG is Brooks fellow and JS
is CITA and PIMS national fellow.  FG was supported in part by
NSF grants AST-0098557 at the University of Washington.


\begin{thebibliography}{44}
\expandafter\ifx\csname natexlab\endcsname\relax\def\natexlab#1{#1}\fi



\bibitem[abadi]{aba}Abadi, M.G., Navarro, J.F., Steinmetz, M. \& Eke, V. 2003 ApJ, 591, 499
\bibitem[athanassoula 1986]{1} Athanassoula, E. \& Sellwood, J. A. 1986, MNRAS, 221, 213
\bibitem[athanassoula 2003]{1} Athanassoula, E, 2003, MNRAS, 341, 1179
\bibitem[Babul, Balogh, Lewis, \& Poole(2002)]{2002MNRAS.330..329B} Babul, 
A., Balogh, M.~L., Lewis, G.~F., \& Poole, G.~B.\ 2002, \mnras, 330, 329 
\bibitem[Balsara (1996)]{1996} Balsara, D.S. 1995, Comp. Phys., 121, 357
\bibitem[Baugh, Cole, \& Frenk(1996)]{1996MNRAS.283.1361B} Baugh, C.~M., 
Cole, S., \& Frenk, C.~S.\ 1996, \mnras, 283, 1361 
\bibitem[Bell, McIntosh, Katz, \& Weinberg(2003)]{2003ApJ...585L.117B} 
Bell, E.~F., McIntosh, D.~H., Katz, N., \& Weinberg, M.~D.\ 2003, \apjl, 
585, L117 
\bibitem[Benson, Bower, Frenk, \& White(2000)]{2000MNRAS.314..557B} Benson, 
A.~J., Bower, R.~G., Frenk, C.~S., \& White, S.~D.~M.\ 2000, \mnras, 314, 
557 
\bibitem[Benz(1990)]{1990nmns.work..269B} Benz, W.\ 1990, Numerical 
Modelling of Nonlinear Stellar Pulsations Problems and Prospects, 269 
\bibitem[Binney, Gerhard \& Silk 2001]{B01} Binney, J., Gerhard, O. \& 
Silk, J., 2001, MNRAS, 321, 471
\bibitem[Bode, Ostriker, \& Turok(2001)]{2001ApJ...556...93B} Bode, P., 
Ostriker, J.~P., \& Turok, N.\ 2001, \apj, 556, 93 
\bibitem[Borgani et al.(2002)]{2002MNRAS.336..409B} Borgani, S., Governato, 
F., Wadsley, J., Menci, N., Tozzi, P., Quinn, T., Stadel, J., \& Lake, G.\ 
2002, \mnras, 336, 409 
\bibitem[Carollo]{} Carollo, M.C., Stiavelli, M., de
  Zeeuw, P.\ T., Seigar, M., \& Dejonghe, H.\ 2001, \apj, 546, 216
\bibitem[cole]{cw} Cole, A.A  \& Weinberg, M.D. 2002, ApJ, 574L, 43
\bibitem[Col{\' i}n, Avila-Reese, \& Valenzuela(2000)]{2000ApJ...542..622C} 
Col{\' i}n, P., Avila-Reese, V., \& Valenzuela, O.\ 2000, \apj, 542, 622 
\bibitem[Colpi, Mayer, \& Governato(1999)]{1999ApJ...525..720C} Colpi, M., 
Mayer, L., \& Governato, F.\ 1999, \apj, 525, 720 
\bibitem[Courteau(1997)]{1997AJ....114.2402C} Courteau, S.\ 1997, \aj, 114, 
2402 
\bibitem[Debattista]{}Debattista, V., Carollo, C.M., Mayer, L., \& Moore, B., to appear on ApJ
\bibitem[Diemand]{2002MNRAS}Diemand, Moore, MNRAS submitted.
\bibitem[Efstathiou, Lake, \& Negroponte(1982)]{1982MNRAS.199.1069E} 
Efstathiou, G., Lake, G., \& Negroponte, J.\ 1982, \mnras, 199, 1069 
\bibitem[Efstathiou et al.(2002)]{2002MNRAS.330L..29E} Efstathiou, G.~et 
al.\ 2002, \mnras, 330, L29 
\bibitem[Eke, Navarro, \& Steinmetz(2001)]{ENS} Eke, V.~R., 
Navarro, J.~F., \& Steinmetz, M.\ 2001, \apj, 554, 114 
\bibitem[Evrard(1988)]{1988MNRAS.235..911E} Evrard, A.~E.\ 1988, \mnras, 
235, 911 
\bibitem[Fall \& Efstathiou(1980)]{1980MNRAS.193..189F} Fall, S.~M.~\& 
Efstathiou, G.\ 1980, \mnras, 193, 189 
\bibitem[Fall83]{fall83} Fall, S.~M.\ 1983, IAU Symp.~100: 
Internal Kinematics and Dynamics of Galaxies, 100, 391 
\bibitem[Fukushige]{2001} Fukushige, T.~\& 
Makino, J.\ 2001, \apj, 557, 533 
\bibitem[Frenk et al.(1999)]{1999ApJ...525..554F} Frenk, C.~S.~et al.\ 
1999, \apj, 525, 554 
\bibitem[Gardner(2001)]{2001ApJ...557..616G} Gardner, J.~P.\ 2001, \apj, 
557, 616 
\bibitem[Ghigna]{2000} Ghigna, S., Moore, B., 
Governato, F., Lake, G., Quinn, T., \& Stadel, J.\ 2000, \apj, 544, 616 
\bibitem[Giovanelli et al.(1997)]{1997AJ....113...53G} Giovanelli, R.,
et al.  1997, \aj, 113, 53 
\bibitem[Haardt \& Madau(1996)]{1996ApJ...461...20H} Haardt, F.~\& Madau, 
P.\ 1996, \apj, 461, 20 
\bibitem[Katz 1992]{nk} Katz, N., 1992, ApJ, 391, 502
\bibitem[Katz \& White(1993)]{1993ApJ...412..455K} Katz, N.~\& White, 
S.~D.~M.\ 1993, \apj, 412, 455 
\bibitem[Klypin, Zhao, \& Somerville(2002)]{2002ApJ...573..597K} Klypin, 
A., Zhao, H., \& Somerville, R.~S.\ 2002, \apj, 573, 597  
\bibitem[Lake \& Carlberg(1988)]{1988AJ.....96.1587L} Lake, G.~\& Carlberg, 
R.~G.\ 1988, \aj, 96, 1587 
\bibitem[Martin(1999)]{1999ApJ...513..156M} Martin, C.~L.\ 1999, \apj, 513, 
156 
\bibitem[Mateo 98]{3}Mateo M. ARAA, 36, 435
\bibitem[Mayer et al.(2001)]{2001ApJ...559..754M} Mayer, L., Governato, F., 
Colpi, M., Moore, B., Quinn, T., Wadsley, J., Stadel, J., \& Lake, G.\ 
2001a, \apj, 559, 754 
\bibitem[Mayer et al.]{Mayer} Mayer, L. \& Wadsley, J. 2004, MNRAS,347, 277
\bibitem{Minniti}{Min} Minniti, D., 1996,  459, 175
\bibitem[Mo, Mao \& White 1998]{MO98} Mo, H . J., Mao, S., \& White, S. D. 
M., 1998, MNRAS, 295, 319
\bibitem[Monaghan(1992)]{1992ARA&A..30..543M} Monaghan, J.~J.\ 1992, \araa, 
30, 543 
\bibitem[Moore et al.(1999)]{1999ApJ...524L..19M} Moore, B., Ghigna, S., 
Governato, F., Lake, G., Quinn, T., Stadel, J., \& Tozzi, P.\ 1999, \apjl, 
524, L19 
\bibitem[Moore et al.(1998)]{1998ApJ...499L...5M} Moore, B., Governato, F., 
Quinn, T., Stadel, J., \& Lake, G.\ 1998, \apjl, 499, L5 
\bibitem[Murali(2000)]{2000ApJ...529L..81M} Murali, C.\ 2000, \apjl, 529, L81 
\bibitem[Navarro \& Steinmetz 2000]{NS00} Navarro J., \& Steinmetz, M.,
2000, ApJ, 538, 477
\bibitem[Navarro \& White(1994)]{1994MNRAS.267..401N} Navarro, J.~F.~\& 
White, S.~D.~M.\ 1994, \mnras, 267, 401
\bibitem[Noguchi]{} Noguchi, M. 2001, \mnras, 328, 353
\bibitem [Pagels]{4} Pagels H. \& Primack, J.R., 1982, Phys. Rev. Lett., 48, 223
\bibitem[Pfenniger \& Friedli(1991)]{1991A&A...252...75P} Pfenniger, D.~\& 
Friedli, D.\ 1991, \aap, 252, 75 
\bibitem[Power et al.]{2002MNRAS} Power, C. et al., 2003 \mnras  338, 14
\bibitem[Perlmutter et al.(1997)]{1997ApJ...483..565P} Perlmutter, S.~et 
al.\ 1997, \apj, 483, 565 
\bibitem[Quinn, Katz, \& Efstathiou(1996)]{1996MNRAS.278L..49Q} Quinn, T., 
Katz, N., \& Efstathiou, G.\ 1996, \mnras, 278, L49 
\bibitem[sembach]{sem} Sembach, K.R. 2003 proceedings of the 2003 STScI May Symposium "The Local Group as an Astrophysical Laboratory".
\bibitem[Somerville(2002)]{2002ApJ...572L..23S} Somerville, R.~S.\ 2002, 
\apjl, 572, L23 
\bibitem[Sommer-Larsen \& Dolgov 1999]{SD99} Sommer-Larson, J., \& Dolgov,
A., astro-ph/9912166,  ApJ in press.
\bibitem[Sommer]{1999ApJ}  Sommer-Larsen, J., Gelato, S., \& Vedel, H.\ 1999, \apj, 519, 501 
\bibitem[Sommer2] {SLP}  Sommer-Larson, J., Gotz, M. \& Portinari, L. 2003 ApJ,  596,47 (S03)
\bibitem[Sparke \& Sellwood(1987)]{1987MNRAS.225..653S} Sparke, L.~S.~\& 
Sellwood, J.~A.\ 1987, \mnras, 225, 653 
\bibitem[Spergel \& Steinhardt(2000)]{2000PhRvL..84.3760S} Spergel, 
D.~N.~\& Steinhardt, P.~J.\ 2000, Physical Review Letters, 84, 3760 
\bibitem[Springel \& Hernquist]{5} Springel, V. \& Hernquist, L. 2002, in press  (astro-ph/0111016) 
\bibitem[Springel \& White(1999)]{1999MNRAS.307..162S} Springel, V.~\& White, S.~D.~M.\ 1999, \mnras, 307, 162 
\bibitem[Stadel]{2002a} Stadel, J., J. Wadsley, and D. C. Richardson 2002. High performance computational astrophysics with \texttt{pkdgrav}/\texttt{gasoline}. In \textit{High Performance Computing Systems and Applications} (N. J.
Dimopoulos and K. F. Lie, Eds.), pp.\ 501--523. Kluwer Academic Publishers, Boston.
\bibitem[tantalo, Chiosi, Bressan]{6}Tantalo R., Chiosi C., Bressan A., 1998. A\&A, 333, 419.
\bibitem[Thacker \& Couchman 2000]{TC99} Thacker, R. J., \& Couchman, H.
M. P., 2000, ApJ, 545, 728 (TC)
\bibitem[Thacker \& Couchman(2001)]{2001ApJ...555L..17T} Thacker, R.~J.~\& Couchman, H.~M.~P.\ 2001, \apjl, 555, L17 (TC01)
\bibitem[Toft, Rasmussen, Sommer-Larsen, \&  Pedersen(2002)]{2002MNRAS.335..799T} Toft, S., Rasmussen, J., 
Sommer-Larsen, J., \& Pedersen, K.\ 2002, \mnras, 335, 799 
\bibitem[Valenzuela \& Klypin(2003)]{2003MNRAS.345..406V} Valenzuela, O.~\& 
Klypin, A.\ 2003, \mnras, 345, 406 
\bibitem[van den Bergh(1996)]{VB} van den Bergh, S. 1996, PASP, 108, 986 
\bibitem[van den Bosch et al 2002]{2002MNRAS submitted} van den
 Bosch, F., Abel,T. Croft, R.A.C., Hernquist,L., Simon D.M. White,
 S.D.M, 2002, ApJ 576, 21
\bibitem[]{7}Wadsley, J., Stadel J. \& Quinn, T. 2003 submitted to New Astronomy.\bibitem[Weinberg \& Katz(2002)]{2002ApJ...580..627W} Weinberg, M.~D.~\& 
Katz, N.\ 2002, \apj, 580, 627 
\bibitem[White \& Rees(1978)]{1978MNRAS.183..341W} White, S.~D.~M.~\& Rees, 
M.~J.\ 1978, \mnras, 183, 341 
\bibitem[Wyse(2002)]{2002b} Wyse, R.F.G. 2002, Euro. Astron. Soc. Pub Ser Vol 2 astro--ph/0204190
\bibitem[Zaldarriaga \& Seljak(2000)]{2000ApJS..129..431Z} Zaldarriaga, 
M.~\& Seljak, U.\ 2000, \apjs, 129, 431 
\bibitem[Zhao, Rich, \& Biello(1996)]{1996ApJ...470..506Z} Zhao, H., Rich, 
R.~M., \& Biello, J.\ 1996, \apj, 470, 506 
\end{thebibliography}
\end{document}